\documentclass[pre,twocolumn,showpacs]{revtex4}
\usepackage{graphicx}
\usepackage{dcolumn}

\begin{document}

\title{Interactions and correlations of particulate inclusions in a columnar phase}

\author{Ronojoy Adhikari}
\email{rjoy@physics.iisc.ernet.in}

\affiliation{Centre for Condensed Matter Theory, Department of Physics,
Indian Institute of Science, Bangalore 560 012, India}

\date{\today}

\begin{abstract} We calculate the elastic field mediated interaction between
macroscopic particles in a columnar hexagonal phase. The interaction is found
to be long-ranged and non-central, with both attractive and repulsive parts. We
show how the interaction modifies the particle correlations and the column
fluctuations. We also calculate the interaction of particles with the
topological defects of the columnar phase. The particle-defect interaction
reduces the mobility of the defects. \end{abstract}


\maketitle

\section{Introduction}

Macroscopic particles suspended in an isotropic fluid - colloidal dispersions -
have been the subject of intensive study for over a century \cite{colloid}.
Recently, a novel class of materials have been realised in which colloidal
particles are dispersed in {\it anisotropic} fluids, typically thermotropic and
lyotropic liquid crystals \cite{poulin,zapot,geetha}. The combination of
anisotropy and long-range order inherent in these mesophases leads to totally
new kinds of interactions between the dispersed particles, which have no
analogues in conventional colloidal
dispersions \cite{brochard,sriram,terentjev,poulinstark,casimir}. 

The most important of these new interactions results from the elastic
deformation induced by a particle in the liquid crystalline host. The overlap
of the the elastic deformations of two particles produces an effective
long-ranged interaction between them. This mechanism, which relies only on the
{\em mean} distortions of the order parameter, implies that other sources of
deformation like topological defects also interact with particles through
long-range forces. Fluctuations of the order parameter together with the
boundary conditions on the particle surface lead to another novel kind of
Casimir-like interaction \cite{kardar}, which can often be important near
special points in the phase diagram. In particle-doped nematics, for
instance, director fluctuations lead to a Casimir force which is stronger than
the van der Waals force near the isotropic-to-nematic transition
 \cite{casimir}.  Interparticle interactions are further complicated by the
inevitable presence of topological defects surrounding the
particles \cite{poulinstark}. These defects compensate the mismatch in the
boundary conditions imposed on the order parameter at the particle surface and
at infinity. To clarify with an example, a particle in a nematic with
director-normal boundary conditions on its surface is equivalent to a radial
hedgehog with unit positive topological ``charge'', and is incompatible with
the requirement of an uniform director configuration at infinity. The mismatch
is relieved by the spontaneous appearance of a defect of unit negative
``charge'', which is tightly bound to the particle.  Such companion defects
usually lead to rapidly decaying, repulsive, power law forces between the
particles \cite{poulinstark}.  The interactions mentioned above, when combined
with the multitude of interactions already present in conventional colloidal
dispersions \cite{colloid}, can drive the particles into spatially organised
structures - chains \cite{poulinstark,cladis}, clusters \cite{raghu}, and even
crystalline arrangements \cite{nazaranko} have been observed for particles in
nematic hosts. 

As a result of these novel interactions, the static and dynamic properties of
(with an obvious extension of terminology) {\em liquid crystalline colloids}
are quite unusual.  The change in the mechanical properties is particularly
dramatic - particles added in  volume fractions as low as several percent  can
totally alter the rheological behaviour of the host phase
 \cite{zapot,geetha,meeker}.  When the particles have a magnetic moment
interesting ferro-liquid crystalline phases are
formed \cite{ramosmag1,ramosmag2}. 

Numerous theoretical studies have been done on these and related effects in
nematic colloids \cite{nemcolloidref}. Comparatively less work has been done on
particle-doped lamellar phases \cite{pershan,turner1,turner2,sens}.  It is
therefore surprising that in spite of recent experiments
 \cite{ramosmag1,ramosmag2}, there has been no theoretical study of
the properties of particle-doped columnar phases. In the following sections
of this paper, therefore, as a first step towards a more detailed theory, we
calculate the elastic field mediated interparticle interactions (section
\ref{effective}), as well as the interactions of particles with point
topological defects (section \ref{topological}).  We conclude the paper with a
critical discussion of our results and possible extensions of the present work.

\section{Particles in a Columnar Hexagonal Phase}
\label{effective}

Let us recall that the columnar phase is macroscopically equivalent to a
regular array of tubes, aligned in, say, the $z$ direction (directions in the
$x$-$y$ plane are denoted as $\perp$) and free to slide along their axes. A
cross-sectional slice normal to the tubes shows a two-dimensional crystal
structure, which, for the case we study, is hexagonal.  Mechanically,
the phase is liquid-like in the $z$ direction, and solid-like and
isotropic in the $\perp$ directions. Elastic deformations are described by a
pair of ``broken-symmetry'' variables ${\bf u}=(u_x,u_y)$ corresponding to the
local displacements of the tubes from their equilibrium position. To lowest
order in gradients, the harmonic free energy density includes contributions from
in-plane bulk and shear deformations, and out-of-plane column
curvatures \cite{degennes}:
\begin{eqnarray} 
\label{hexf}
{\cal F}_{hex}&=&{B\over2}(\nabla_{\perp}\cdot{\bf
u})^2+{K_3\over2}\left(\nabla_z^2 {\bf u}\right)^2
\nonumber\\
&+&{C\over2}[(\nabla_x u_x -\nabla_y u_y)^2 +(\nabla_y u_x + \nabla_x u_y)^2].
\end{eqnarray} 
Here $B$ and $C$ are first-order elastic constants (with dimensions of energy
density) corresponding to bulk and shear deformations respectively, while $K_3$
is a second-order elastic constant (with the dimension of force) corresponding
to column curvature. An elastic description of this sort is valid at length
scales large compared to the column spacing $d$. The colloidal particles are
modelled as monodisperse spheres of radius $a$ and are described by a
concentration field $c({\bf r})=\sum_{\nu}\delta({\bf r-r}_{\nu})$, where ${\bf
r}_{\nu}$ is the center-of-mass coordinate of the $\nu$th particle. A
description of this sort is valid at length  scales large compared to
the particle radius $a$, when in effect, the particles can be treated as
point masses. In this dilute limit, where the typical distance between particles
is much larger than their size, all short-range interactions can be ignored,
and the particles need to be assigned only an ideal gas entropy.  As in the
theory of point defects in three dimensional crystals, it is assumed that the
particles behave as force dipoles (``centers of dilation or
compression'') \cite{eshelby} and produce a body force proportional to $\nabla
c({\bf r})$ at the point $({\bf x}_{\perp},z)={\bf r}$.  This is equivalent to
a coupling of the form $c({\bf r})E({\bf u})$ in the free energy, where $E({\bf
u})=\nabla_{\perp}\cdot{\bf u}$ is the local dilation. It is not difficult to
see that to  linear order in ${\bf u}$ and lowest order in gradients, this is
the only possible symmetry-allowed coupling of the concentration to the column
displacement \cite{note1}. With these considerations, then, the free energy
density (in units of $k_BT$) is
\begin{equation} 
\label{fullf}
{\cal F}(c,{\bf u}) = {\cal
F}_{hex}+c({\bf r})\left[ln\frac{c({\bf r})}{c_0}-1\right]+ \alpha c({\bf
r})E({\bf u}). 
\end{equation} 
The second term in (\ref{fullf}) accounts for the entropy of the particles
whose mean concentration is $c_0$. $\alpha$ is a coupling constant,which on
dimensional grounds must be of order $Ba^3$. Note that in our units $B$ has the
dimensions of inverse volume and that $\alpha$ is dimensionless. The
microscopic details of the boundary conditions on the particle surface are
incorporated phenomenologically in the sign and magnitude of $\alpha$.  For
positive $\alpha$ the columns are repelled away from the particle surface
leading to a local dilation; this situation is most appropriate for inert
colloidal particles, whose size $a$ is appreciably larger than the column
spacing $d$. For negative $\alpha$ the columns are are attracted towards the
particle leading to a local compression; this situation may be appropriate for
small biological inclusions, which are known to exert local inward forces on
their surroundings \cite{turner1,safran}.  Henceforth, we shall assume $\alpha$
to be positive. It is worth emphasising that our theory is valid for
separations large compared to {\em both} $d$ and $a$, which represent,
respectively, the scales beyond which the elastic and concentration field
descriptions are meaningful.  

\subsection{Effective Interactions}

Integrating out the displacement field ${\bf u}$ from $F[c,{\bf u}]=\int
d^3{\bf r}{\cal F}(c,{\bf u})$ we obtain an effective free energy 
\begin{equation}
\label{concdef}  
F_c=-ln\left(\int{\cal D}{\bf u}e^{-F[c,{\bf u}]}\right)
\end{equation} 
for the concentration fluctuations. Since our interest lies only in the mean
displacements of the columns, the Gaussian integral in
(\ref{concdef}) can instead be replaced by the minimum  of $F[c,{\bf u}]$ over
${\bf u}$.  Setting the variation $\delta F/\delta{\bf u}$ to zero, we obtain a
pair of coupled linear differential equations for the components $(u_x,u_y)$.
Transforming to Fourier space (ie multiplying by $e^{-i{\bf q\cdot r}}$ and
integrating over $d^{3}{\bf r}$), and solving the resulting algebraic equation
for the Fourier components, we have 
\begin{equation}
\label{green1}
{\bf u_q}=\frac{i\alpha{\bf q_{\perp}}}{(B+C)q_{\perp}^{2}+K_3q_z^{4}}c_{\bf q}.
\end{equation}
This equation gives the (mean) displacement field for an imposed concentration
fluctuation $c_{\bf q}$. Using it to eliminate the displacement field from (\ref{fullf}), we obtain
the effective free energy as 
\begin{equation}
\label{eff1}
F_c =\int d^3{\bf r}\ c({\bf r})\left[ln\frac{c({\bf
r})}{c_0}-1\right]+\frac{1}{2}\int_{\bf q}c_{\bf q}V({\bf q})c_{\bf -q}
\end{equation}
where 
\begin{equation}
V({\bf q})=\frac{-\alpha^{2}q_{\perp}^{2}}{(B+C)q_{\perp}^{2}+K_3q_z^{4}}
\end{equation}
and we have used the shorthand $\int_{\bf q}\equiv\int\frac{d^3{\bf
q}}{(2\pi)^3}$. The second term in (\ref{eff1}) represents the effective
interaction mediated by the column displacements. The kernel $V({\bf q})$ can
be identified with the Fourier transform of the interparticle potential $V({\bf
r})$, with the understanding that the correct physics is captured only for
lengths large compared to the particle size.  With that in mind, the interparticle 
potential is given by 
\begin{eqnarray} 
\label{pot1}
V({\bf r})&=&{\alpha^2\lambda_3^2\over K_3}\nabla_{\perp}^{2}G({\bf x_{\perp}},z) 
\nonumber\\
&=&{\alpha^2\lambda_3^2\over K_3}\nabla_{\perp}^{2}\int_0^{\infty} dt
\frac{e^{-z^2/4\lambda_3 x_{\perp} \cosh t}}{(\lambda_3 x_{\perp}\cosh
t)^{1/2}}
\end{eqnarray}
where $\lambda_3^2=K_3/(B+C)$. The Fourier transform $G({\bf x_{\perp}},z)$ of
the kernel $(q_{\perp}^2+\lambda_3^2q_z^4)^{-1}$ can be obtained explicitly in
terms of modified Bessel functions \cite{prost}, but we prefer the integral
representation given above.

\begin{figure}
\includegraphics[width=7cm,height=7cm]{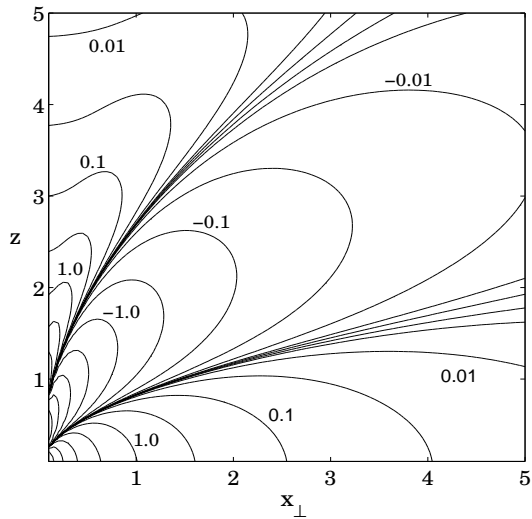}
\caption {\label{cplot1}Plot of the equipotentials of $V({\bf
r})\sim\nabla^2_{\perp} G({\bf r})$. The coordinates are in units of
$\lambda_3$.The contour lines have been drawn on equal logarithmically spaced
intervals from .001 to 1000, and -0.001 to -1000. The potential is even in $z$.}
\end{figure}

The potential $V({\bf r})$ presents several interesting features. It is
non-central (not a function of $r$ only) and radially non-monotonic. This last
property prevents us from inferring the sign of the radial component of the
force from that of the potential. Therefore, the attractive or repulsive
character of the interaction cannot be correlated with the sign of the
potential, and the forces have to be considered explicitly. However, in the $z$
and $\perp$ directions the potential {\em is} monotonic and decays with a power
law.  Near the $z$-axis (for $z^2\gg4\lambda_3 x_{\perp}$) the potential
vanishes as $|z|^{-5}$, leading to a rapidly damped and repulsive force along
the columns.Near the $x$-$y$ plane (for $z^2\ll4\lambda_3 x_{\perp}$) the
potential decays more slowly as $x_{\perp}^{-5/2}$, implying a repulsive force
in the $\perp$ direction.  In a general direction the nature of the interaction
is best visualised through the plot of the equipotentials of $V({\bf r})$ shown
in Fig.\ref{cplot1}.  From the plot it is clear that the radial component of
the force is positive near the $z$-axis and the $x$-$y$ plane, and negative in
the region bounded by the two paraboloids $z^2\sim\lambda x_{\perp}$. Broadly
speaking, the force felt by a particle at $({\bf x}_{\perp},z)$ due to another
particle at the origin varies from repulsive, to attractive, to repulsive again
as the polar angle $\theta=\tan^{-1}({z/x_{\perp}})$ varies from $0$ to
$\pi/2$. Since the potential is even in $z$, the sequence repeats as the angle
decreases from $\pi$ to $\pi/2$. A qualitatively similar angular dependence is
also found in the quadrupolar forces between particles in a
nematic \cite{sriram}; the distance dependence, though, is totally different. 

Given the nature of the interparticle forces described above, we might expect
an attraction induced aggregation of the particles.  However, repulsive forces
due to the companion topological defects mentioned in the introduction may
dominate at short distances, and arrest the aggregation. In that case, it is
likely that particles will spontaneously organise in chains oblique to the
columns. This process will be dominated by Brownian motion, and not the elastic
interaction, which though long-ranged, decays too rapidly to affect the
kinetics of aggregation.   

The correction to the interparticle potential due to a finite density of
particles may be calculated quite easily within a mean field approximation
 \cite{cl}.  Augmenting the free energy $F_c$  with an external potential
$V_{ext}({\bf r})$ which couples to the concentration, the equilibrium
concentration is obtained by demanding $\delta F_c/\delta c=0$. This gives 
\begin{equation}
c({\bf r})=c_{0}\exp\left[-V_{ext}({\bf r})-\int d^3{\bf r'}c({\bf r'})V({\bf
r-r'})\right].
\end{equation}
The mean (or effective) potential is defined by $c({\bf r})=c_{0}e^{-U({\bf r})}$.
Using this to eliminate the density from the previous equation 
and identifying the external potential with that of a fixed particle at the origin,
we obtain the following self-consistent equation for $U({\bf r})$:
\begin{equation}
U({\bf r})=V({\bf r})+c_0\int d^3{\bf r'} V({\bf
r-r'})\exp\left[-U({\bf r'})\right].
\end{equation}
Assuming that $U\ll1$ we can linearise the above integral equation. Neglecting
the term corresponding to the background concentration and Fourier transforming
we find
\begin{equation}
U({\bf q})=\frac{V({\bf q})}{1+c_0V({\bf q})}=\frac{-\alpha^{2}q_{\perp}^{2}}
{(B+C-c_0\alpha^2)q_{\perp}^{2} +K_3q_z^{4}}.
\end{equation}
The mean potential is of the same functional form as the "bare" interparticle
potential, but with a ``renormalised'' characteristic length
$\lambda_3^2(\alpha)= K_3/(B+C-c_0\alpha^2)$. Recalling that a similar
calculation with a Coulomb potential $V({\bf q})\sim q^{-2}$ leads to charge
screening \cite{cl}, we must conclude that, despite the long-ranged
interaction, many-particle effects are quite modest in our case.

\subsection{Correlation Functions}

In the absence of  interactions the particles in our model have only
configurational entropy and their two-point correlations are like those of an
ideal gas: $<\delta c({\bf r})\delta c({\bf r'})>\sim\delta({\bf r-r'})$. However,
elastic interactions lead to quite complex correlations, as we
show below. Defining the dimensionless structure factor $S(\bf q)$ as the
Fourier transform of ${1\over c_0}<\delta c({\bf r})\delta c({\bf r'})>$ we
find from the effective free energy in (\ref{eff1}) that
\begin{eqnarray}
S({\bf q})&=&\frac{1}{1+ c_0V({\bf q})}
\nonumber\\
&=&1+\frac{c_0\alpha^{2}q_{\perp}^{2}}{(B+C-c_0\alpha^2)q_{\perp}^{2}+K_3q_z^{4}}.
\end{eqnarray}
The Fourier transform of the pair correlation function $h({\bf q})=S({\bf
q})-1$ is proportional to $-U({\bf q})$, and therefore the pair correlations
themselves are well described by the contour plot of Fig.\ref{cplot1} with all
the signs reversed.  The correlations are negative along the $z$ and $\perp$
axes, indicating a reduction in the particle concentration compared to the mean
value of $c_0$, while in the region bounded by the paraboloids
$z^2\sim\lambda_3x_{\perp}$ the correlations are positive, indicating an
enhancement of the concentration. This is broadly consistent with our earlier
picture of particle chains oblique to the columns. It may be possible to
measure the particle structure factor by small-angle X-ray scattering, and the
form of $S({\bf q})$ above would then provide a test of the particle-column
coupling assumed in our theory. 

While the column displacements induce particle correlations, the particle
aggregates in turn modify the column fluctuations.  In particle-doped
lamellar phases, this feedback mechanism has been  shown to change the usual
Caill\'e \cite{caille} form of the structure factor \cite{sens}.
To calculate the column displacement correlations in the presence of particles,
we first define the displacement field such that $\langle\nabla_i u_j\rangle=0$
for a mean concentration $c_0$. This requires that we modify the
particle-column coupling to $\delta c({\bf r})E({\bf u})$.  Next, we define an
effective free energy $F_{\bf u}$ for the column fluctuations (in a way similar
to the definition of $F_c$) by integrating out the concentration field.
Carrying out the integration by steepest descent we obtain the 
effective free energy 
\begin{equation} 
\label{eff2}
F_{\bf u} =\int d^3{\bf r}\left[{\cal F}_{hex} -c_0\left(\alpha\nabla_{\perp}
\cdot{\bf u} -e^{-\alpha\nabla_{\perp}\cdot{\bf u}}\right)\right].
\end{equation} 
This shows that the coefficients of all symmetry-allowed terms which contain
$\nabla_{\perp}\cdot{\bf u}$ are ``renormalised'' by the presence of particles.
Within our harmonic description the only relevant term, obtained by expanding
the exponential term to quadratic order, is $\alpha^2(\nabla_{\perp}\cdot{\bf u})^2$. 
Retaining only this term in (\ref{eff2})  and defining the longitudinal and 
transverse parts of ${\bf u_q}$ as $q^2{\bf u}_L={\bf q}({\bf q\cdot u_q})$
and ${\bf u}_T={\bf u_q}-{\bf u}_L$ we obtain the $u_{L(T)}-u_{L(T)}$
correlations as
\begin{equation} 
|{\bf u}_T|^2=\frac{1}{Cq_{\perp}^{2} +K_3q_z^{4}},
\end{equation} 
\begin{equation} 
|{\bf u}_L|^2=\frac{1}{(B+C-c_0\alpha^2)q_{\perp}^{2} +K_3q_z^{4}}.
\end{equation} 
The transverse correlations are identical to those of the pure columnar phase,
which indicates that there are no corrections to $C$ and $K_3$ due to the
particles.  The longitudinal correlations while identical in form, have a
modified bulk modulus $B_{c}= B-c_0\alpha^{2}$. Particles thus produce a
softening of the columnar phase.  

\section{Particle-defect interactions}
\label{topological}
\subsection{Point defects in the columnar phase}

Topological defects are common in liquid crystalline phases, being easily
produced under conditions of processing such as shear flows.Much of the
interesting macroscopic behaviour of liquid crystalline colloids comes from the
interaction of the colloidal particles with the topological defects of the host
 \cite{zapot,geetha,meeker}. For example, the solid-like response of
particle-doped cholesteric \cite{zapot} and smectic \cite{geetha} phases is due
to the formation of a stable network of defects. The colloidal particles play
the crucial role of {\em stabilising} this network, which otherwise would
anneal away. It is likely that elastic interactions do not play a dominant role
in such processes, and that short-range forces, not captured in a
coarse-grained description such as ours, are more important. Nevertheless,
elastic interactions between particles and topological defects are known to
have important consequences \cite{kosevich}, and so in what follows, we
calculate this interaction for a point defect configuration in the columnar
phase. The statistical mechanics of thermally activated point defects
 \cite{jain,nelson} may be modified in the presence of particles and a knowledge
of particle-defect interactions could be potentially useful in a study of this
phenomenon. An interesting duality (see below) between the columnar and
lamellar phases has been noted by de Gennes and Prost \cite{degennes} and
explored in detail by Selinger and Bruinsma \cite{bruinsma}. This suggests that
it may be of interest to compare particle-defect interactions in the lamellar
and columnar phases and we conclude this section with such a comparison. 

\begin{figure}
\includegraphics[width=5cm,height=5cm]{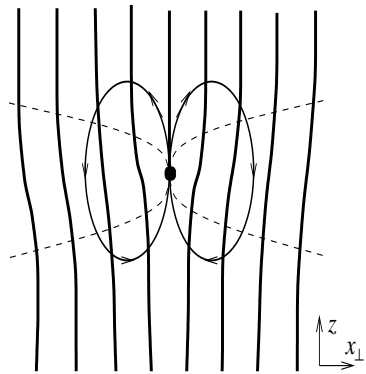}
\caption{
\label{hexdefect}
A ``tail'' defect configuration of topological charge $m=+1$ 
with the thick lines showing the column deformations column near
the defect. The particle-defect potential vanishes on the paraboloids
$z^2\sim\lambda_3 x_{\perp}$ (dashed lines), beyond which  it is mainly
repulsive for $z>0$ and attractive for $z<0$. The thin lines show typical
trajectories of overdamped particles: they are expelled from the regions
of dilation ($z>0$) and attracted to the regions of compression ($z<0$). 
}
\end{figure}

The defect configuration we consider is a single {\em column ending}
 \cite{prost,bruinsma}, illustrated in Fig. \ref{hexdefect}.  This is not unduly
restrictive, since any linear defect which consists of column endings (eg. a
transverse edge dislocation) can be constructed out of a collection of such
point defects \cite{prost}. The ease with which a column can break depends on
what constitutes the ``column'', and therefore, on what the particular
realisation of the columnar phase is.  In polymeric columnar phases, the
columns consist of long chains arranged end-to-end along their length.  Here,
thermally activated column scission can occur only at those places where one
end of a chain meets that of another. (Polymeric liquid crystals are found
predominantly in the nematic and isotropic phases; however, longer and stiffer
biopolymers like DNA can also positionally order into columnar phases
 \cite{polymerhex}).  Since columns in discotic and micellar phases are composed
of weakly bonded molecules, thermally activated scission can take place 
anywhere along the length of the columns. In all of these examples, the column
scission energy is finite and thus a state of thermal equilibrium at non-zero
temperature will always contains a finite number of such defects. 
The ends of a scissioned column may either separate along the $z$-axis to form
a string of vacancies, or may translate past each other in the $\perp$
direction to form a string of interstitials \cite{jain}.  Such strings will
begin from and end on the column extremities, the lower of which, by
convention, is called a ``tail''(Fig. \ref{hexdefect}) and the upper a
``head'' \cite{jain}.  At high temperatures the entropy of wandering of the
string can overcome the energy due to its line tension, leading to a string
proliferation, and producing isolated ``heads'' or
``tails'' \cite{prost,frey}. Signatures of such defects have been observed in
recent experiments \cite{endpoint}.  An isolated semi-infinite string running
along the $z$-axis with its terminus at the origin imposes the constraint
that \cite{bruinsma} 

\begin{equation}
\nabla_{\perp}\cdot{\bf u}=m{\delta(\bf x_{\perp})\over\rho_0}\theta(z),
\end{equation}
where the topological ``charge'' $m$ is $+1$ ($-1$) for a tail (head), and
$\rho_0$ is the {\em areal} density of the columns.  Clearly this is nothing
but a line of centers of dilation and is easily treated in the framework of the
previous section. Formally, a column  terminus at the origin is identical to a
line of point particles distributed with the concentration 
\begin{equation}
c({\bf r})=m{(B+C)\over\alpha\rho_0}\delta({\bf x_{\perp}})\theta(z).
\end{equation}
Combining this with the expression in (\ref{green1}) for the displacement
produced by a distribution of particles, the displacement due to a column ending 
is obtained as \cite{note2}  
\begin{equation}
{\bf u_q}={i{\bf q_{\perp}}\over q_{\perp}^{2}+\lambda_3^2q_z^{4}}
{m\over\rho_0 iq_z}.
\end{equation}
The displacement for a collection of column endings distributed with the
topological ``charge'' density $m({\bf r})$ follows from the superposition of the
individual displacements. Stated mathematically, this means that the total
displacement is  
\begin{equation}
\label{green2}
{\bf u_q}={i{\bf q_{\perp}}\over q_{\perp}^{2}+\lambda_3^2q_z^{4}}
{m_{\bf q}\over\rho_0 iq_z},
\end{equation}
with $m_{\bf q}$ being the Fourier transform of $m({\bf r})$.  To calculate the
particle-defect interaction, the displacement field is decomposed as ${\bf
u_{\bf q}}={\bf u_q}(c_{{\bf q}})+{\bf u_q}(m_{{\bf q}})$, where the first part
(due to the particles) is obtained from (\ref{green1}) and the second part (due to
the defects) is obtained from (\ref{green2}).  The free energy then separates
out as $F=F_c + F_{mm} + F_{cm}$, with

\begin{figure}
\includegraphics[width=7cm,height=7cm]{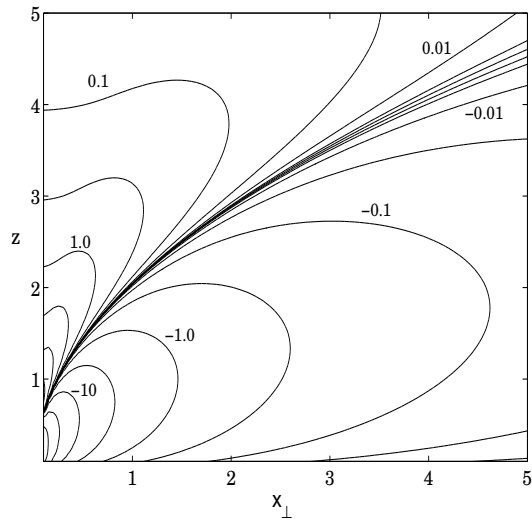}
\caption {\label{cplot2}Plot of the equipotentials of $V_{cm}({\bf
r})\sim\nabla^3_{z}G({\bf r})$. The coordinates are in units of
$\lambda_3$.The contour lines have been drawn on equal logarithmically spaced
intervals from .001 to 100, and -0.001 to -100. The potential is odd in $z$.}
\end{figure}

\begin{equation}
\label{mm}
F_{mm}=\frac{1}{2}\int_{\bf q}\frac{q_{\perp}^{2}}{\rho_0^2q_z^2}
\frac{m_{\bf q}m_{\bf-q}}{q_{\perp}^{2}+\lambda_3^2q_z^{4}}{K_3\over\lambda_3^2}
\end{equation}
and
\begin{equation}
\label{cm}
F_{cm}=\int_{\bf q}\frac{i\alpha q_{\perp}^{2}}{\rho_0q_z}
\frac{m_{\bf q}c_{\bf-q}}{q_{\perp}^{2}+\lambda_3^2q_z^{4}}.
\end{equation}
The Fourier transforms of the defect-defect and particle-defect interactions
can now be read off as the kernels of $m_{\bf q}m_{\bf-q}$ and $c_{\bf q}m_{\bf-q}$
in (\ref{mm}) and (\ref{cm}) respectively. The transformation to real space can
be simplified by noting that  
\begin{equation} 
\nabla_{\perp}^2G({\bf x_{\perp}},z)=\lambda_3^2\nabla_{z}^4G({\bf
x_{\perp}},z). 
\end{equation} 
Using this property, $q_{\perp}^2$ can be replaced  by $\lambda_3^2q_z^4$
in the {\em numerators} of the kernels in (\ref{mm}) and (\ref{cm}).
Alternatively, a conjugate Green's function can be introduced as in
 \cite{bruinsma} and \cite{nelson}. With either of the methods, the final form
of the interaction kernels turn out to be
\begin{equation}
V_{mm}({\bf q})={K_3\over\rho_0^2}{q_z^2\over q_{\perp}^2+\lambda_3^2q_z^4},
\end{equation}
and
\begin{equation}
V_{cm}({\bf q})={\alpha\lambda_3^2\over\rho_0}
{iq_z^3\over q_{\perp}^2+\lambda_3^2q_z^4}.
\end{equation} 
Apart from the line tension terms which we ignore, the defect-defect
interaction is identical to the result of \cite{bruinsma} and has been
discussed in detail there \cite{note3}. The particle-defect interaction can be
represented in real space as
\begin{equation} V_{cm}({\bf r})={m\alpha\lambda_3^2\over\rho_0}
\nabla_{z}^{3}G({\bf x_{\perp}},z).
\end{equation}

As might have been anticipated, the particle-defect potential is non-central
and radially non-monotonic, and is {\em odd} in the $z$-coordinate.  The
behaviour in the $z$ and $\perp$ directions is simple. Along the columns (for
$z^2\gg4\lambda_3 x_{\perp}$) the potential decays as sgn($z$)$|z|^{-4}$
indicating a repulsive force for $z>0$ and an attractive force for $z<0$. In
the $x$-$y$ plane, the $\perp$ component of the force vanishes by symmetry and
the $z$ component goes to zero as $x_{\perp}^{-5/2}$.  The detailed form of the
interaction for positive $z$ is shown in the equipotential plot of
Fig.\ref{cplot2}. The general conclusion is that independent of the sign of the
defect, the particles are attracted to the regions of compression, and repelled
away from the regions of dilation that the defect {\em alone} would have
produced, see Fig.\ref{hexdefect}. Driven by the attractive component of the
interaction, and aided by Brownian motion, a cloud of particles should collect
around a column ending,  a situation reminiscent of the Cottrell clouds of
vacancies and interstitials around dislocations in metals \cite{kosevich}.
Normally, such clouds decrease the dislocation mobility \cite{kosevich} and the
same can be assumed to happen here. Since defect motion is one of the main
mechanisms of stress relaxation in ordered phases, less mobile defect strings
may lead to an increase in the viscosity of the columnar phase. 

It is important to note that expression (\ref{green2}) can be used to obtain
the interaction of a particle with {\em any} linear defect which can be
represented as a collection of column endings. For example, the interaction of
a transverse edge dislocation with a particle can be obtained from
(\ref{green2}) by setting $m({\bf r})\sim\delta(y)\delta(z)$ and repeating the
steps leading to (\ref{cm}).

\subsection{Line defects in the lamellar phase} 

There is an interesting line-plane duality \cite{degennes,bruinsma} between
defect-free columnar and lamellar phases which is apparent in a pictorial
representation: a set of lines representing the columns (aligned, say, along
the $z$-axis) when rotated by $\pi/2$ can equally well be viewed as a set of
planes representing the lamellar layers (now stacked normal to the $z$-axis).
Mechanically, the lamellar phase is a 2d liquid and a 1d solid, while the
columnar phase is a 1d liquid and a 2d solid. Though this duality is lost in
the presence of an arbitrary ensemble of defects (for example screw
dislocations), nonetheless there is one important exception. It is easy to see
from Fig. \ref{hexdefect} that an endpoint defect (the terminus of a line) in
the columnar phase corresponds to an edge dislocation (the terminus of a plane)
in the lamellar phase. Going beyond pictures, there are close connections in
the energetics of these ``dual''defects, which have been explored in
 \cite{bruinsma}. This prompts us to ask if similar connections may be found in
the interactions of endpoints and edge dislocations with particles. To answer
this question, we first examine the particle-dislocation interaction for an
arbitrary distribution of dislocations, and then specialise to the case of an
edge dislocation.  

The free energy density of a particle-doped lamellar phase where the particles 
couple to the local layer dilation has been obtained in \cite{turner1}. Excluding the
entropic terms, in our notation it is 
\begin{equation}
\label{eff3}
F={\bar B\over2}\int d^3{\bf r}\left[(\nabla_z u)^2 + \lambda^2(\nabla_{\perp}^2 u)^2
+\bar\alpha c({\bf r})\nabla_z u\right],
\end{equation}
where $u$ is the local displacement of the layers from their equilibrium
position, $B$ is the bulk modulus, $K=B\lambda^2$ is a second order elastic
constant corresponding to layer curvature, and the positive constant
$\bar\alpha$ is the strength of the particle-layer coupling. In the presence of
a dislocation density ${\bf b(r)}$ the distortion ${\bf v}=\nabla u$ acquires a
singular transverse part determined by \cite{pershan}
\begin{equation}
\nabla\times {\bf v=b(r)}.
\end{equation}
The total distortion in Fourier space is given by \cite{cl}
\begin{equation}
\label{dist1}
{\bf v_q}({\bf b_q})=\frac{i{\bf q}\times{\bf b_q}}{q^2}+
i{\bf q}h_{\bf q},
\end{equation}
where $h_{\bf q}$, the longitudinal part of the distortion, is obtained from the 
equation of equilibrium ${\delta F/\delta u}=0$. The solution 
\begin{equation}
\label{eff4}
h_{\bf q}=\frac{\bar\alpha iq_zc_{\bf q} }{q_z^2 + \lambda^2 q_{\perp}^4} 
-\frac{q_z(1-\lambda^2q_{\perp}^2)}{q_z^2 + \lambda^2 q_{\perp}^4}
\frac{\left({\bf\hat z\cdot q\times b_q}\right)}{q^2}
\end{equation}
is a sum of the particle and defect contributions. When the distortion in
(\ref{dist1}) is inserted into the (Fourier transformed) free energy
(\ref{eff4}), the latter separates into three pieces: the first contains only
the dislocation density \cite{cl}, the second only the concentrations
 \cite{turner1}, and the third is a cross-term which represents the
particle-dislocation interaction. Displaying only this last term, we have
\begin{equation}
\label{eff5}
F_{cb}={\bar B}\int_{\bf q} \frac{\bar\alpha\lambda^2q_{\perp}^2}
{q_z^2 + \lambda^2q_{\perp}^4}({\bf \hat z\cdot q\times b_q})c_{-\bf q}.
\end{equation}
This gives the interaction energy of an arbitrary distribution of particles
with an arbitrary number of dislocations. Two important special cases are
linear edge and screw dislocation of infinite extent.  The dislocation density
for a screw dislocation of infinite extent and Burgers vector $b{\bf\hat z}$ is
${\bf b(r)}=b\delta(x)\delta(y){\bf \hat z}$, which implies that ${\bf
b_q}=2\pi b\delta(q_z){\bf \hat z}$. Inserting this into (\ref{eff5}) we find
that there is no {\em elastic} interaction between a particle and the screw
component of a dislocation within a linear elastic theory.  For the case of
interest, an infinite edge dislocation running along the $y$-axis with Burgers
vector $b{\bf\hat z}$, ${\bf b(r)}= b\delta(x)\delta(z){\bf \hat y}$, and ${\bf
b_q}=2\pi b\delta(q_z){\bf \hat z}$, and so together with (\ref{eff5})  the
particle-edge dislocation interaction kernel is obtained as
\begin{equation}
V_{cb}({\bf q})=\frac{\bar B\bar\alpha \lambda^2 q_{\perp}^2}
{q_z^2 + \lambda^2q_{\perp}^4} 2\pi biq_x\delta(q_y).
\end{equation}
Transforming to real space, the interaction potential is
\begin{equation}
V_{cb}({\bf r})=\frac{\bar B\bar\alpha b }{8}\frac{x}{\sqrt{\pi\lambda|z|^3}}
\exp\left({-x^2\over4\lambda|z|}\right).
\end{equation}

\begin{figure}
\includegraphics[angle=90,width=5cm,height=5cm]{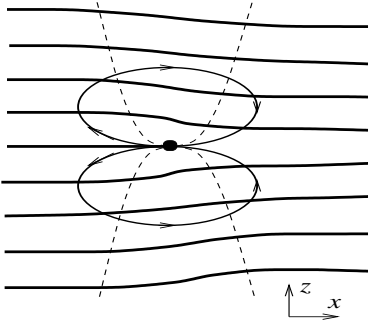}
\caption{
\label{lamdefect}
An infinite edge dislocation of Burgers charge $b=+1$ running along the
$y$-axis with the thick lines showing the layer deformations near the defect.
The dashed lines show the parabolas $x^2=4\lambda|z|$ for arbitrary values of
$y$.  The thin lines show typical trajectories of overdamped particles: they
are expelled from the regions of dilation ($x<0$) and attracted to the regions
of compression ($x>0$).  Notice that this configuration is obtained by rotating
Fig.\ref{hexdefect} through $\pi/2$. 
}
\end{figure}

The interaction is long-ranged along the $z$-axis (for $x^2\ll4\lambda|z|$) and
exponentially damped along $x$-axis (for $x^2\gg4\lambda|z|$). By symmetry, the
$z$-component of the force vanishes on the $z$-axis, while both the $x$ and $z$
components are exponentially small near the $x$-axis.  Irrespective of the sign
of $b$, the particles are attracted to the compression regions that the
dislocation alone would have produced, and repelled away from the dilation
regions, see Fig.(\ref{lamdefect}). To compare this interaction with its dual
in the columnar phase, we first note that the interaction is characterised by
the parabolas $x^2=4\lambda|z|$ which occur frequently in problems related to
dislocations in lamellar phases \cite{degennes}. The particle-defect potential
in the columnar phase is characterised instead by the paraboloids
$z^2\sim\lambda_3 x_{\perp}$.  In both cases the maximum elastic deformation is
confined to the regions within these surfaces, compare Fig.(\ref{hexdefect})
and Fig.(\ref{lamdefect}).  The force is long-ranged along the solid-like
directions (the $\perp$ direction in the columnar phase, and the $z$ direction
in the lamellar phase) and short-ranged in the liquid-like directions. There is
no interaction between a particle and an edge dislocation when both have the
same $z$ coordinate, a feature also seen in the interaction between two edge
dislocations \cite{degennes}.  In contrast, the particle-defect interaction in
the columnar phase is strongly repulsive in the $z$ direction. The formation of
Cottrell clouds, and the consequent reduction in the dislocation mobility
should be expected in the lamellar phase as well.

\section{Conclusion} 

To summarise, we have calculated the elastic interactions of colloidal
particles in a columnar hexagonal phase by modelling them as centers of
dilation. The interaction is long-ranged and non-central, and has both
attractive and repulsive parts. We have shown that this interaction leads to
correlations between the particle positions, and to a reduction in the bulk
modulus $B$ of the columnar phase.  We have also shown how particles interact
elastically with defects in columnar and lamellar phases. In both phases,
particles aggregate around the defects driven by the attractive component of
the interaction and reduce the defect mobility. One likely consequence of this
is an increase in the viscosity of the phase. 

We mention below several of the directions in which the present work may be
extended.  First, the elastic interactions of particles of more general shapes
like cylinders and discs should be examined.  Such inclusions can no longer be
modelled as force dipoles, and force distributions containing higher multipoles
will be needed \cite{eshelby}. The interaction of plate-like particles would
have a direct relevance to the physics of clay-block copolymer
composites \cite{groenewold}. Second, the effects of boundaries should be
investigated. Free boundaries require the introduction of image sources, which
will have long-range interactions with particles in the bulk. Thus, boundary
effects can be expected to strongly influence all processes in the bulk, and in
particular, the formation of particle aggregates. Third, the nature of the
topological defects surrounding a particle should be examined.  Our
coarse-grained description cannot {\em predict}  the defect configuration
(though useful information can be gained if we {\em assume} it) and a more
microscopic description would be better suited for the purpose.

As a final, somewhat speculative remark,  we should indicate that the present
calculation may be of some relevance in biological systems where inclusions may
be present in tissues of columnar hexagonal structure \cite{bouligand}.
Biological inclusions generate local inward forces \cite{turner1,safran}, and a
minimal description of such inclusions could be in terms of the centers of
compression (corresponding to a negative $\alpha$) in our theory. Similar
elastic interactions between proteins in a lamellar phase \cite{turner2}, and
cells in an isotropically elastic tissue have already been
investigated \cite{safran}.

\acknowledgements 

The author would like to thank Professor Sriram Ramaswamy for suggesting the
problem, useful discussions, and a critical reading of the manuscript, and
Professor Chandan Dasgupta for helpful suggestions. Partial support from
Hindustan Lever Limited is gratefully acknowledged. 




\begin{references}
\bibitem{note2} Prost \cite{prost} has obtained a similar result by treating the
column ending as a screw dislocation dipole.
\bibitem{note3}Note that the authors' interpretation of the sign of the
defect-defect potential is not correct. See the discussion following our equation 
(\ref{pot1}).
\end{references}
\end{document}